\documentclass[aps,prb,preprint,superscriptaddress,colorlinks=true,linkcolor=black,urlcolor=black,citecolor=black,anchorcolor=black,colorlinks,bookmarks=false,citecolor=blue,linkcolor=black,urlcolor=blue]{revtex4-1}

\usepackage{lineno}

\usepackage{ulem}
\usepackage{bm,color,amsmath,amssymb,latexsym,graphicx,psfrag,accents,float}
\usepackage{amscd,dsfont,wasysym,mathrsfs,mathtools}
\usepackage{multirow}
\usepackage{dcolumn}
\usepackage{xcolor}
\usepackage{comment}
\usepackage{booktabs} 
\usepackage{upgreek}
\usepackage[utf8]{inputenc}
\usepackage{braket}
\usepackage{times}
\usepackage{amsfonts}
\usepackage{mathrsfs}
\usepackage{graphicx}
\usepackage{dcolumn}
\usepackage{bm}
\usepackage{color}

\usepackage[colorlinks,bookmarks=false,citecolor=blue,linkcolor=red,urlcolor=blue]{hyperref}
\usepackage{amsmath}
\usepackage{ulem}
\begin{document}
\renewcommand{\theequation}{S\arabic{equation}}
	\newcommand{\beginMethods}{%
		\setcounter{table}{0}
		\renewcommand{\thetable}{S\arabic{table}}%
		\setcounter{figure}{0}
		\renewcommand{\thefigure}{S\arabic{figure}}%
	}
\newcommand{\bk}{{\bf k}}
\newcommand{\bB}{{\bf B}}
\newcommand{\bv}{{\bf v}}

\title{Distinct switching of chiral transport in the kagome metals KV$_3$Sb$_5$ and CsV$_3$Sb$_5$}
\author{Chunyu Guo${}^{\dagger}$}\affiliation{Max Planck Institute for the Structure and Dynamics of Matter, Hamburg, Germany}
\author{Maarten R. van Delft}\affiliation{High Field Magnet Laboratory (HFML - EMFL), Radboud University, Toernooiveld 7, 6525 ED Nijmegen, The Netherlands}
\author{Martin Gutierrez-Amigo}\affiliation{Centro de Física de Materiales (CSIC-UPV/EHU), Donostia-San Sebastian, Spain}
\affiliation{Department of Physics, University of the Basque Country (UPV/EHU), Bilbao, Spain}
\author{Dong Chen}\affiliation{Max Planck Institute for Chemical Physics of Solids, Dresden, Germany}\affiliation{College of Physics, Qingdao University, Qingdao, China}
\author{Carsten Putzke${}^{}$}
\affiliation{Max Planck Institute for the Structure and Dynamics of Matter, Hamburg, Germany}
\author{Glenn Wagner}\affiliation{Department of Physics, University of Zürich, Zürich, Switzerland}
\author{Mark H. Fischer}\affiliation{Department of Physics, University of Zürich, Zürich, Switzerland}
\author{Titus Neupert}\affiliation{Department of Physics, University of Zürich, Zürich, Switzerland}
\author{Ion Errea}\affiliation{Centro de Física de Materiales (CSIC-UPV/EHU), Donostia-San Sebastian, Spain}
\affiliation{Donostia International Physics Center, Donostia-San Sebastian, Spain}
\affiliation{Fisika Aplikatua Saila, Gipuzkoako Ingeniaritza Eskola, University of the Basque Country (UPV/EHU), Donostia-San Sebastian, Spain}
\author{Maia G. Vergniory}\affiliation{Max Planck Institute for Chemical Physics of Solids, Dresden, Germany}
\affiliation{Donostia International Physics Center, Donostia-San Sebastian, Spain}
\author{Steffen Wiedmann}\affiliation{High Field Magnet Laboratory (HFML - EMFL), Radboud University, Toernooiveld 7, 6525 ED Nijmegen, The Netherlands}
\affiliation{Radboud University, Institute for Molecules and Materials, Nijmegen 6525 AJ, Netherlands}
\author{Claudia Felser}\affiliation{Max Planck Institute for Chemical Physics of Solids, Dresden, Germany}
\author{Philip J. W. Moll${}^{\dagger}$}\affiliation{Max Planck Institute for the Structure and Dynamics of Matter, Hamburg, Germany}

\date{\today}
\maketitle
\normalsize{$^\dagger$Corresponding authors: chunyu.guo@mpsd.mpg.de(C.G.);
philip.moll@mpsd.mpg.de(P.J.W.M.).}

\section{abstract}

The kagome metals AV$_3$Sb$_5$ (A=K,Rb,Cs) present an ideal sandbox to study the interrelation between multiple coexisting correlated phases such as charge order and superconductivity. So far, no consensus on the microscopic nature of these states has been reached as the proposals struggle to explain all their exotic physical properties. Among these, field-switchable electric magneto-chiral anisotropy (eMChA) in CsV$_3$Sb$_5$ provides intriguing evidence for a rewindable electronic chirality, yet the other family members have not been likewise investigated. Here, we present a comparative study of magneto-chiral transport between CsV$_3$Sb$_5$ and KV$_3$Sb$_5$. Despite their similar electronic structure, KV$_3$Sb$_5$ displays negligible eMChA, if any, and with no field switchability. This is in stark contrast to the non-saturating eMChA in CsV$_3$Sb$_5$ even in high fields up to 35~T. In light of their similar band structures, the stark difference in eMChA suggests its origin in the correlated states. Clearly, the V kagome nets alone are not sufficient to describe the physics and the interactions with their environment are crucial in determining the nature of their low-temperature state.

\section{introduction}

Metals hosting kagome nets have recently proven to be a fruitful avenue to explore correlated topological materials\cite{kagomeReview,syozi1951statistics,kang2020topological,kang2020,Yin2020,Yin2018,Ye2018,Howard2021,ortiz2019new}. Their orbital and magnetic frustration generically gives rise to Dirac points and potentially flat bands, which are associated with non-trivial behavior such as giant intrinsic anomalous Hall effect\cite{Yin2020,Ye2018} and topologically protected boundary states\cite{Yin2020,Howard2021}. The recently reported family AV$_3$Sb$_5$ (A = K, Rb, Cs) presents an intriguing example of electronic instabilities on a kagome lattice driven by strong correlations\cite{ortiz2019new,ortiz2020cs,Kang2022,TitusAdd,Neupert2022,guo2023correlated}. They all jointly undergo a charge order transition distorting the kagome lattice at $T_{CDW} \sim 100$ K. The main open question in this field concerns the types of broken symmetries within that ordered state, most prominently the fate of time-reversal and mirror symmetries. While a mirror symmetric structure appears in X-Ray diffraction\cite{CVS_Xray}, experimental evidence for broken symmetries mounts, including electronic C2 anisotropy\cite{xiang2021twofold,Nie2022,Nema2}, a chiral charge-density-wave state observed in STM experiments\cite{zhao2021cascade,Kchiral} and three-state nematicity in the optical Kerr effect\cite{Xu2022}. A further direct consequence of broken mirror symmetries in electric conductors is a current-direction dependent voltage response called electrical magneto-chiral anisotropy (eMChA)\cite{TeChiral,BiHelix,carbon,TTlO4,CrNbSB,MnSi}. This has been found recently in CsV$_3$Sb$_5$, which further demonstrates an electronic chirality within the charge order\cite{Guo2022}. The direction of chiral transport is uniquely switchable by a magnetic field, which points to its origin in a mirror-symmetry-breaking correlated state rather than the common structural chirality found in diodes. Clearly, further experiments probing this dichotomy between electronic and structural chirality are called for. 

 A first key step concerns the generality of magneto-chiral transport among the AV$_3$Sb$_5$ family of compounds. As an iso-structural analog to CsV$_3$Sb$_5$, KV$_3$Sb$_5$ displays a similar charge order formation at high temperature as well as a superconducting ground state\cite{ortiz2021superconductivity}. Based on this similarity, its electronic chirality has also been explored. A rotational symmetry breaking chiral charge order is consistently observed yet contradictory conclusions have been made about whether its chirality can be controlled by the magnetic field\cite{Kchiral,li2022rotation}. Here, we examine the magneto-chiral transport properties of KV$_3$Sb$_5$ with a side-by-side comparison to CsV$_3$Sb$_5$, offering a great opportunity for exploring the critical factors for the magneto-chiral transport among the AV$_3$Sb$_5$ series of compounds and beyond. 



\subsection*{Transport properties and electronic fermiology}
Experiments probing eMChA in CsV$_3$Sb$_5$ have shown it to be extremely susceptible to external perturbations such as strain or magnetic field\cite{Guo2022,guo2023correlated}. A quantitative comparison of different AV$_3$Sb$_5$ compounds requires to significantly reduce the uniaxial strain due to thermal contraction difference. We therefore decouple the crystalline sample mechanically as much as possible from the substrate by fabricating lithographic springs that act as ultra-soft mechanical support and electric contacts\cite{Guo2022}. The central crystalline microstructure has been carved from a single crystal using focused-ion-beam (FIB) milling (Fig. 1). It features a Hall-bar device with six electric terminals. One of the current leads is fixed directly to the Si-substrate via FIB-assisted Pt-deposition to reduce the torque distortion at high magnetic field, and the other five electric contacts are supported only by soft gold-coated membrane springs (100~nm SiN$_x$ and 150~nm Au). This has been previously shown to reduce the forces on similar microstructures to below 50 bar \cite{Guo2022}, a key prerequisite to observe eMChA in them.  

With these low-strain devices, we firstly explore the electronic transport properties of both compounds within the linear response regime which sets the basis for non-linear chiral transport. A clear anomaly in the temperature dependence of resistivity reveals the CDW transition temperatures at 94 K and 76 K for CsV$_3$Sb$_5$ and KV$_3$Sb$_5$ respectively, consistent with previous reports\cite{ortiz2019new,ortiz2020cs} (Fig. 1). The smaller lattice constant in KV$_3$Sb$_5$ compared to CsV$_3$Sb$_5$ could be considered as a positive chemical pressure effect, and this decrease of $T_{CDW}$ indeed matches expectations from hydrostatic pressure experiments\cite{du2021pressure,zhang2021pressure,PressureAM}. However, the reported superconducting transition is suppressed down to $T_c~\approx~0.7~$K in KV$_3$Sb$_5$, which is significantly lower than simple hydrostatic pressure arguments may explain\cite{ortiz2021superconductivity}. Meanwhile it displays a broader charge density wave transition with a less pronounced jump of resistivity, as well as a moderately larger residual resistivity at base temperature. It has been reported that $T_c$ and K vacancies are closely related in this compound\cite{ortiz2021superconductivity}, which implies lattice defects as a possible origin for the lower $T_c$ and the increased residual resistivity in KV$_3$Sb$_5$.  

To explore more quantitatively the differences between KV$_3$Sb$_5$ and CsV$_3$Sb$_5$, we turn towards their electronic band structure studied both theoretically and experimentally. Ab-initio band structure calculations\cite{QE-2017,perdew1996generalized,DALCORSO2014337} have been performed without taking the charge-order formation into account (Fig. 2). Since both KV$_3$Sb$_5$ and CsV$_3$Sb$_5$ crystallize in the P6/mmm space group with the kagome net formed by the V-atoms, it is not surprising that they share a clear similarity and slightly differ due to the unit cell change and the opening of the gap at the M point. This gap results in a reconstruction of the M pockets, as depicted in the Fermi Surface insets of Fig. 2. To confirm this similarity in the band structures, we have also examined its validity experimentally. The fermiology of KV$_3$Sb$_5$ has rarely been studied in detail\cite{KVS1QO,AHE_KVS}, unlike CsV$_3$Sb$_5$\cite{fu2021quantum,SAxy,CVS-thin,ortiz2021fermi,MB-CVS,CVS-interlayer,huang2022mixed}. To directly contrast the Fermi surfaces of these two compounds, we have performed magneto-transport measurements of the membrane-based device elongated along the c-axis up to 35 T with a rotation of field direction from c to a'-axis for both materials (see supplement). A third-order polynomial fit for the field-dependence of magnetoresistance allows us to extract the Shubnikov-de-Haas oscillations. 

The quantum oscillation frequencies disperse in lockstep for both compounds, directly evidencing the similarity of their electronic structures. Multiple orbits are detected that show 2D and 3D characteristics. Several low frequency oscillations below 500 T consistently appear with field applied almost within the kagome plane (see supplement), demonstrating the 3D nature of the corresponding Fermi surfaces, consistent with previous reports\cite{huang2022mixed}. The main high frequencies observed can be divided into two branches around 1700 T and 700 T. These frequencies are comparable with the Fermi surfaces located around A and H points obtained from ab-initio calculations\cite{ortiz2021fermi}, while the Brillouin-zone-sized pocket around $\Gamma$ point is not observed in our measurements. The angular dependences of the frequencies follow nicely the general description of a 2D Fermi surface ($F$ $\propto$ 1/cos($\theta$)), suggesting the quasi-2D nature of these Fermi surfaces. The similarity in fermiology between these compounds results in the consistent electronic properties among the AV$_3$Sb$_5$ family such as the previously proposed orbital loop current and correlated charge order. In light of this similarity, the striking difference in eMChA between the compounds is puzzling, as will be shown next. 


\subsection*{Significant suppression of eMChA in KV$_3$Sb$_5$}

The electrical magneto-chiral anisotropy, eMChA, can occur in the absence of mirror symmetries in the system. It results in a polarity-dependent resistance value as $R(\boldsymbol{B},\boldsymbol{I}) \neq R(\boldsymbol{B},-\boldsymbol{I})$, which is usually detected by the second-harmonic voltage generation with low-frequency AC currents\cite{TeChiral,BiHelix,carbon,TTlO4,CrNbSB,MnSi}. Recently, eMChA with a field-switchable forward direction has been reported\cite{Guo2022}, pointing to spontaneous mirror symmetry breaking of the correlated order and setting the system apart from other structurally chiral conductors, in which the handedness is firmly imprinted during materials synthesis. Indeed, the sign of eMChA is controlled by a small out-of-plane field component $B_c$, demonstrating a field-switchable electronic chirality. 

To further explore and compare the electronic chirality in CsV$_3$Sb$_5$ and KV$_3$Sb$_5$, we have performed measurements of second harmonic voltage generation due to eMChA for both compounds up to 35 T (Fig. 3). A non-saturating, field-asymmetric V$_{2\omega}$ signal is observed in CsV$_3$Sb$_5$, which increases beyond 4 $\mu$V at $B$ = 35 T. Due to the equally non-saturating magnetoresistance, the second harmonic voltage displays a nearly $B^3$ dependence throughout the entire field window, suggesting the electronic chirality is not affected by the in-plane magnetic field. As the lowest-order coupling between magnetic field and current, the chiral contribution to electrical conductance $\Delta\sigma$ is proportional to $V_{2\omega}~/~V_{\omega}^2$ and displays a linear field-dependence. This naturally explains the nearly $B^3$-dependence of $V_{2\omega}$\cite{Guo2022}. Most importantly, the sign of $\Delta\sigma$ is reversed when the magnetic field rotates across the kagome plane, suggesting a direct correspondence between the handedness of electronic chirality to the direction of the out-of-plane field component. Even in fields of 35 T, no saturation to the fast growth of the eMChA signal is observed, indicating that any putative crossover occurs at yet higher fields.

In absence of saturation, the sizable chiral contribution $\Delta$R/R reaches 1.2\% at 35 T. This value is substantial for a second-order correction term. It is clear, however, that this growth cannot continue much further. The eMChA signal is already appreciable, yet with a continued growth following a B$^3$ dependence, it may overtake the resistance itself, when $\Delta R/R$ = 1. A naive extrapolation places this transition at 150~T, and further high-field investigations of eMChA may be successful at detecting the incipient deviations.

On the contrary, KV$_3$Sb$_5$ displays an almost negligible second-harmonic signal. At the same current density as CsV$_3$Sb$_5$, the second harmonic voltage stays below 30 nV up to 35 T, more than two orders of magnitude smaller. Furthermore, the sign of $\Delta \sigma$ cannot be switched by tilting the magnetic field through the kagome planes as in CsV$_3$Sb$_5$ (Fig. 3). On both sides of the planes ($\theta$ = -1 and 2), the sign of the signal remains unchanged. As the second harmonic signal in KV$_3$Sb$_5$ is so small, it is experimentally difficult to determine if at all any non-trivial eMChA exists in it. The main difficulty comes from asymmetric Joule heating due to accidentally imbalanced magnetoresistances at the contacts. The striking difference between these materials is clear already from the raw data. 

\section{Discussion and Outlook}
It is difficult to reconcile the stark difference between their eMChA signals with the similarities of their single-particle spectrum, hence likely the key differences reside in interacting physics and differences in the electronic order. One structural difference is the higher density of vacancies in KV$_3$Sb$_5$ compared to CsV$_3$Sb$_5$. This reflects in the higher residual resistivity, the lower magnetoresistance, comparatively weaker SdH oscillations and a reduced transport anisotropy at zero field. This scenario finds support in the angular dependence of the magnetoresistance. In CsV$_3$Sb$_5$, a pronounced spike is observed with field applied within the kagome plane. This is a signature of coherent interlayer transport as despite the emergence of small 3D pockets due to charge-order formation\cite{huang2022mixed}, the Brillouin zone is still predominantly occupied by quasi-2D Fermi surfaces at low temperature. However this spike is strongly suppressed in KV$_3$Sb$_5$, consistent with enhanced decoherence scattering. 

Depending on the origin of eMChA, such enhanced scattering in KV$_3$Sb$_5$ influences the system in several possible ways. Firstly, since the temperature coefficients for transport remains metallic in all directions and the residual resistivity does not differ significantly (only by a factor of 2), it is hard to explain the huge difference in eMChA just by the smearing effect due to increased isotropic, achiral scattering sites. This suggests that the defining factor of eMChA in KV$_3$Sb$_5$ may reside beyond just the band structure effect within the (chiral) ordered phase. Secondly, if eMChA originates in the scattering on chiral domains, the disorders/vacancies distort the kagome net formed by the V-atoms and can act as the pinning centers that imprint the electronic chiral domains to a fixed pattern. Since these point disorders are achiral, this fixed pattern is naturally balanced in chirality. Moreover, this fixed pattern is stable against the out-of-plane magnetic field and therefore does not have a minority/majority chirality. This means the chiral scattering process is always canceled out which results in the strong suppression of eMChA. Last but not least, the possibility of an achiral bulk state cannot be ruled out. Despite the report of chiral or even switchable chiral state by STM measurements\cite{zhao2021cascade,Kchiral}, the possibility still exists that such a state only appears at the surface. Previous studies demonstrate that CsV$_3$Sb$_5$ is located at a tipping point between different correlated orders and the subtle differences in electronic structures we observed could drive KV$_3$Sb$_5$ sufficiently deep in an achiral state, eliminating the origin of chiral transport observed in CsV$_3$Sb$_5$. Other origins of the possible achiral bulk state in KV$_3$Sb$_5$, such as the potential difference in phonon spectrum, should also be further examined.


All proposals suggest that the surprising suppression of eMChA in KV$_3$Sb$_5$ relies on subtle electronic features. Therefore to differentiate these scenarios and identify the origin of the strong eMChA signal in CsV$_3$Sb$_5$ it is of particular interest to revisit eMChA in AV$_3$Sb$_5$ at slightly different aspects. The first thing to establish is the detailed relation between defect concentration and the strength of eMChA. This can be achieved via controlling the effective chemical substitution such as Sn-doping\cite{Sn-doping}, K-vacancies\cite{ortiz2021superconductivity} or electron radiation\cite{PDO-ele,he-osc}. If the K-vacancies can be reduced to a level, where the difference in scattering rate between KV$_3$Sb$_5$ and CsV$_3$Sb$_5$ becomes negligible, one can explore the possible intrinsic difference in (switchable) electronic chirality between them. Furthermore, for systematic doping studies, if the amplitude of eMChA is directly proportional to the defect concentration, the single-particle scenario is valid and the chiral transport is swamped by the increase of isotropic, achiral scattering sites. On the other hand if the eMChA is dramatically suppressed only at a threshold of doping level, this would suggest that once a sufficient number of pinning centers is formed, the domain pattern is locked and therefore eMChA vanishes. Based on this scenario, it is also worth exploring whether a stronger magnetic field can overcome the pinning energy of the locked domain pattern near the critical doping level which provides further evidence for the chiral domain scenario. 

In summary, we have reported a distinct switching of chiral transport in the kagome metal KV$_3$Sb$_5$ and CsV$_3$Sb$_5$. KV$_3$Sb$_5$ displays a negligible electronic chiral transport signature compared to CsV$_3$Sb$_5$. Moreover, the direction of the chiral transport is no longer switchable by the magnetic field. The minor difference in electronic structure between these compounds apparently contrasts strongly with the massive difference in magneto-chiral transport. This is clearly beyond the simple description on the single-particle level, where the electronic correlation becomes significant. These results point towards exotic correlated states with extreme tunability/sensitivity in AV$_3$Sb$_5$ compounds, calling for further attention.

\clearpage

\noindent \textbf{Acknowledgements: } This work was funded by the European Research Council (ERC) under the European Union’s Horizon 2020 research and innovation programme (MiTopMat - grant agreement No. 715730 and PARATOP - grant agreement No. 757867). This project received funding by the Swiss National Science Foundation (Grants  No. PP00P2\_176789). This work was supported by HFML-RU/NWO-I, member of the European Magnetic Field Laboratory (EMFL). M.G.V., I. E. and M.G.A. acknowledge the Spanish Ministerio de Ciencia e Innovacion  (grant PID2019-109905GB-C21). M.G.V., C.F., and T.N. acknowledge support from FOR 5249 (QUAST) lead by the Deutsche Forschungsgemeinschaft (DFG, German Research Foundation). M.G.V. acknowledges partial support to European Research Council grant agreement no. 101020833. This work has been supported in part by Basque Government grant IT979-16. This work was also supported by the European Research Council Advanced Grant (No. 742068) “TOPMAT”, the Deutsche Forschungsgemeinschaft (Project-ID No. 247310070) “SFB 1143”, and the DFG through the W\"{u}rzburg-Dresden Cluster of Excellence on Complexity and Topology in Quantum Matter ct.qmat (EXC 2147, Project-ID No. 390858490).

\noindent \textbf{Author Contributions:} Crystals were synthesized and characterized by D.C. and C.F.. The experiment design, FIB microstructuring and high-field magnetotransport measurements were performed by C.G., M.R.V.D., S.W., C.P., and P.J.W.M.. Band structures were calculated by M.G.A., I.E. and M.G.V.. G.W., M.H.F. and T.N. developed the general theoretical framework for eMChA analysis, and the analysis of experimental results has been done by C.G., C.P. and P.J.W.M.. All authors were involved in writing the paper.

\noindent \textbf{Competing Interests:} The authors declare that they have no competing financial interests.\\

\noindent \textbf{Data Availability:} Data that support the findings of this study will be deposited to Zenodo with the access link displayed here.

\section*{References}

\clearpage

\clearpage

\begin{figure}
	\centering
\includegraphics[width = 0.9\linewidth]{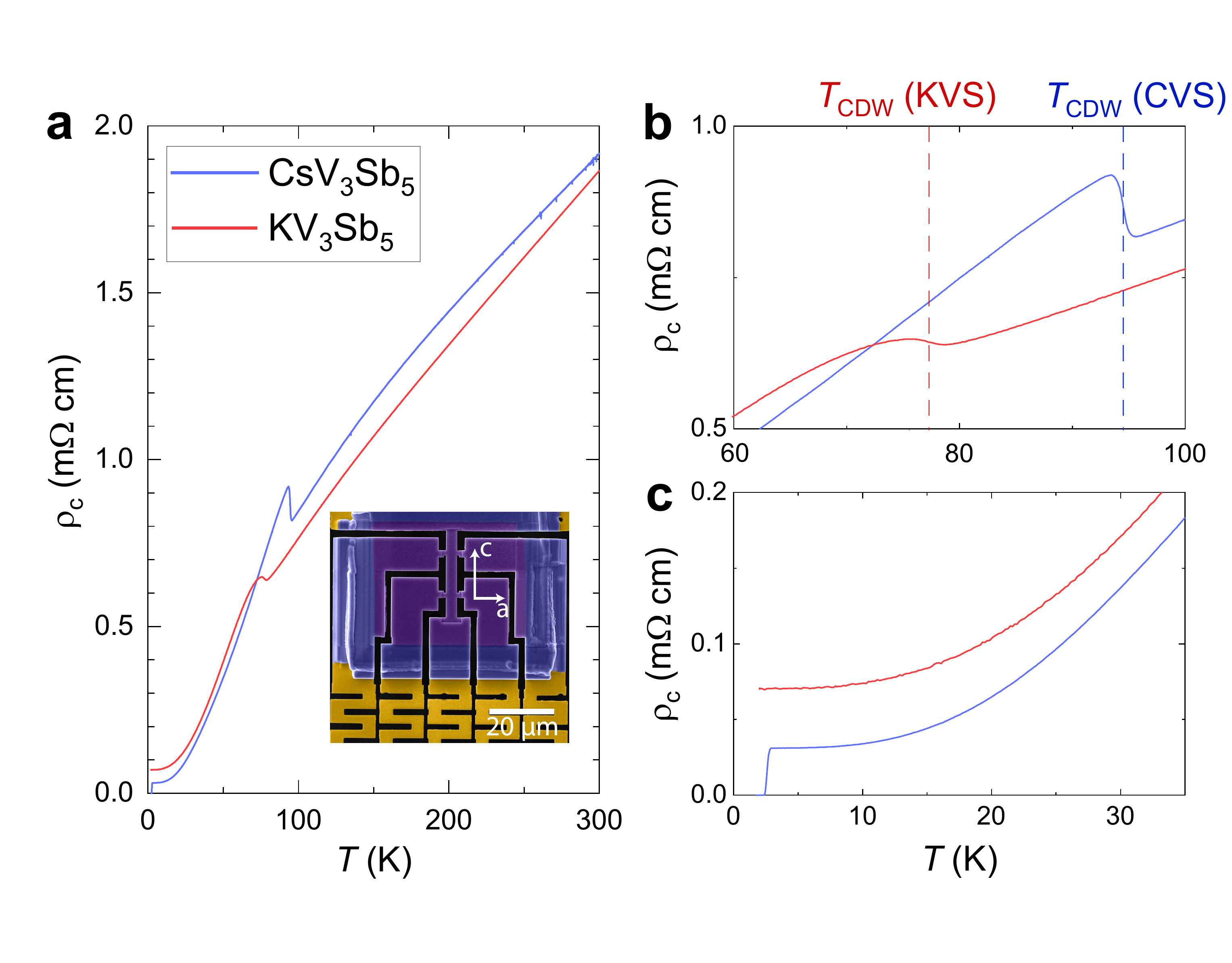}
		\caption{\textbf{Temperature-dependence of resistivity.} (a) Temperature dependence of electric resistivity for CsV$_3$Sb$_5$ and KV$_3$Sb$_5$. The inset displays the membrane-based microstructure which features a Hall-bar geometry with long-axis along the c-direction. (b) Both materials display a clear resistivity jump due to the charge-density-wave (CDW) transition. The transition temperature $T_{CDW}$ is 94 K and 76 K for CsV$_3$Sb$_5$ and KV$_3$Sb$_5$ respectively. (c) Low temperature resistivity for both Cs and K compounds. No superconducting transition is found in KV$_3$Sb$_5$ down to $T$ = 2~K, and its residual resistivity is larger compared to CsV$_3$Sb$_5$.}
	\label{Tdep}
\end{figure}
\clearpage

\begin{figure}
	\centering
\includegraphics[width = 0.9\linewidth]{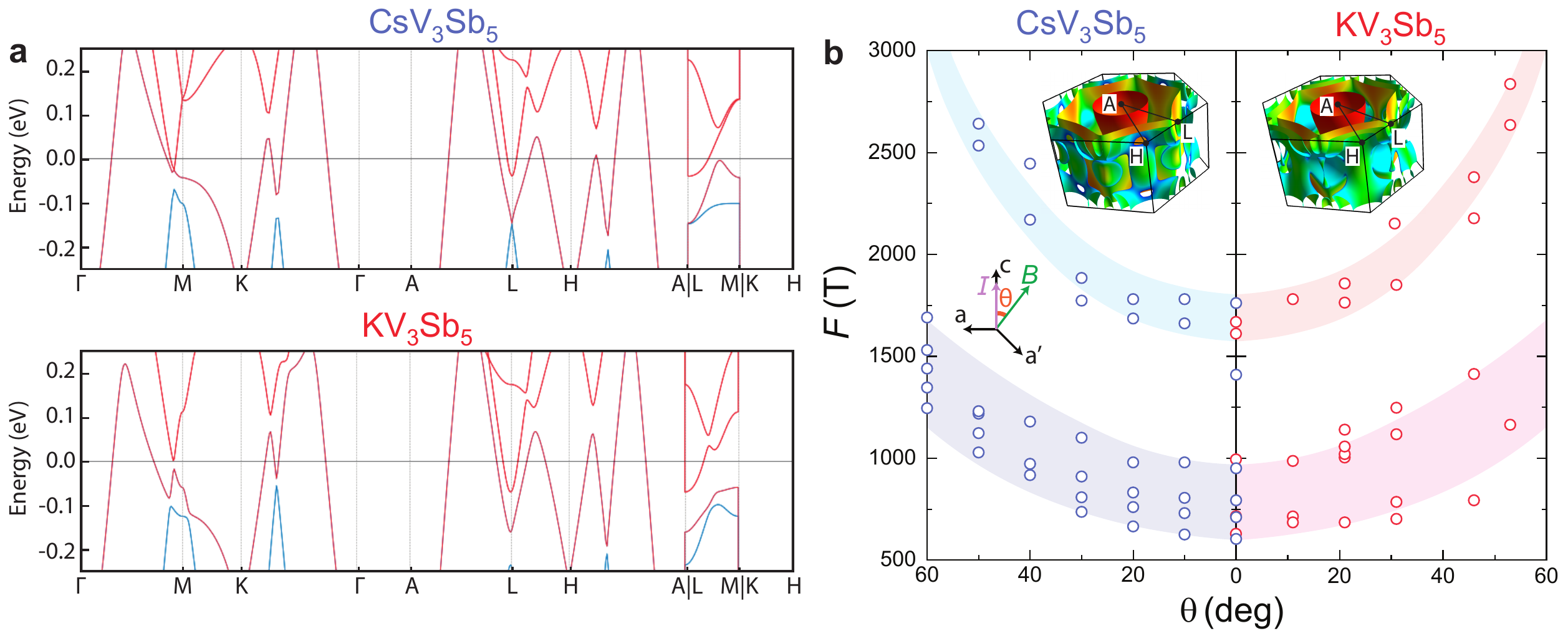}
		\caption{\textbf{Electronic structure and fermiology.} (a) Electronic band structure of both CsV$_3$Sb$_5$ and KV$_3$Sb$_5$ calculated by density-functional theory (DFT). (b) Angular dependence of quantum oscillation frequency. A side-by-side comparison between CsV$_3$Sb$_5$ and KV$_3$Sb$_5$ suggest a qualitative similarity in fermiology, as also demonstrated by the DFT calculation displayed in the inset.}
	\label{Cs135}
\end{figure}
\clearpage

\begin{figure}
	\centering
\includegraphics[width = 0.9\linewidth]{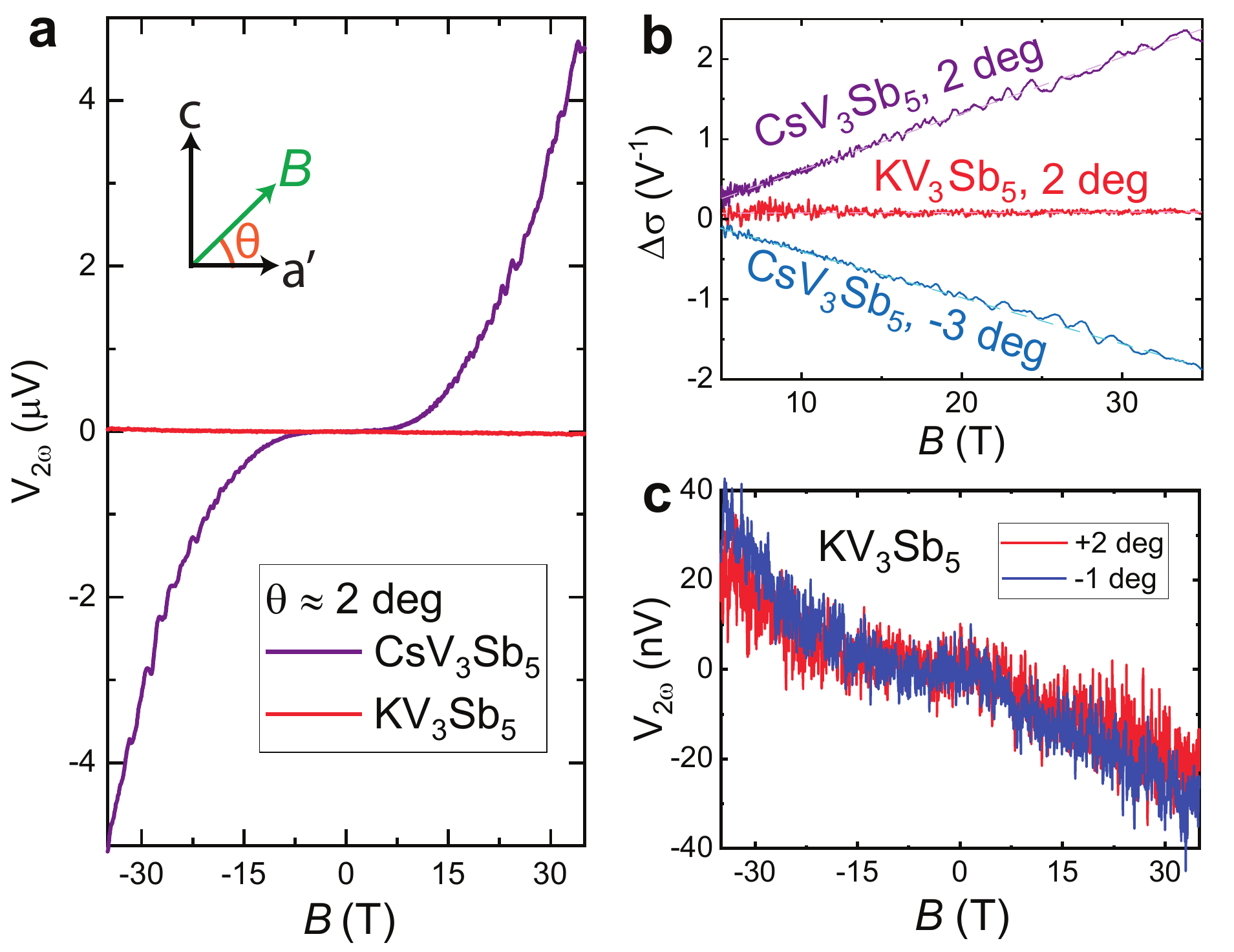}
		\caption{\textbf{Suppression of eMChA in KV$_3$Sb$_5$ and comparison to CsV$_3$Sb$_5$.} (a) Field dependence of second harmonic voltage. A clear B$^3$-dependence is observed in CsV$_3$Sb$_5$ as consistent with previous report\cite{Guo2022}. The magneto-chiral conductivity displays a clear field-linear dependence as shown in (b), indicating robust magneto-chiral transport up to 35 T. On the other hand, the second harmonic voltage measured in KV$_3$Sb$_5$ is about two orders of magnitude smaller compared to CsV$_3$Sb$_5$. Moreover, this tiny signal remains nearly unchanged with magnetic field rotated across the kagome plane (c), suggesting its non-switchable nature that is distinct from CsV$_3$Sb$_5$.}
	\label{Main}
\end{figure}
\clearpage

\begin{figure}
	\centering
\includegraphics[width = 0.9\linewidth]{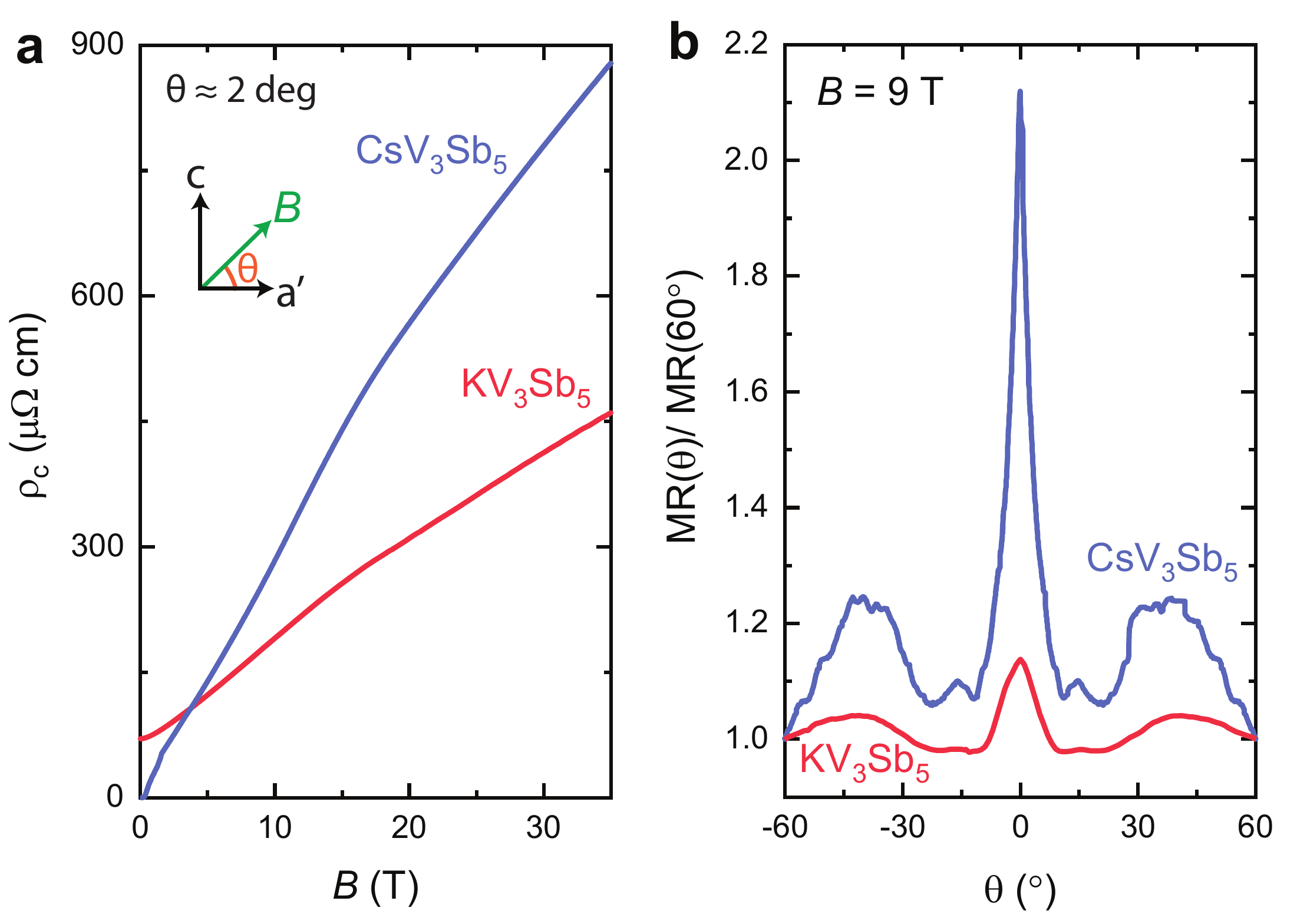}
		\caption{\textbf{In-plane spike of magnetoresistance.} (a) The nearly in-plane magnetoresistance for KV$_3$Sb$_5$ compared to CsV$_3$Sb$_5$. (b) Angular dependence of magnetoresistance further demonstrates the difference between the Cs- and K-compounds. A significant spike can be observed for CsV$_3$Sb$_5$ when the magnetic field is applied within the kagome plane. On the other hand, the magnetoresistance is also maximized with the same field configuration for KV$_3$Sb$_5$, yet the magnitude of the spike is strongly reduced.}
	\label{Theory}
\end{figure}
\clearpage

\end{document}


\renewcommand{\theequation}{S\arabic{equation}}
	\newcommand{\beginsupplement}{%
		\setcounter{table}{0}
		\renewcommand{\thetable}{S\arabic{table}}%
		\setcounter{figure}{0}
		\renewcommand{\thefigure}{S\arabic{figure}}%
	}
\title{Supplementary materials for "Distinct switching of chiral transport in the kagome metals KV$_3$Sb$_5$ and CsV$_3$Sb$_5$"} 

\author{Chunyu Guo${}^{\dagger}$}\affiliation{Max Planck Institute for the Structure and Dynamics of Matter, Hamburg, Germany}
\author{Maarten R. van Delft}\affiliation{High Field Magnet Laboratory (HFML - EMFL), Radboud University, Toernooiveld 7, 6525 ED Nijmegen, The Netherlands}
\author{Martin Gutierrez-Amigo}\affiliation{Centro de Física de Materiales (CSIC-UPV/EHU), Donostia-San Sebastian, Spain}
\affiliation{Department of Physics, University of the Basque Country (UPV/EHU), Bilbao, Spain}
\author{Dong Chen}\affiliation{Max Planck Institute for Chemical Physics of Solids, Dresden, Germany}\affiliation{College of Physics, Qingdao University, Qingdao, China}
\author{Carsten Putzke${}^{}$}
\affiliation{Max Planck Institute for the Structure and Dynamics of Matter, Hamburg, Germany}
\author{Glenn Wagner}\affiliation{Department of Physics, University of Zürich, Zürich, Switzerland}
\author{Mark H. Fischer}\affiliation{Department of Physics, University of Zürich, Zürich, Switzerland}
\author{Titus Neupert}\affiliation{Department of Physics, University of Zürich, Zürich, Switzerland}
\author{Ion Errea}\affiliation{Centro de Física de Materiales (CSIC-UPV/EHU), Donostia-San Sebastian, Spain}
\affiliation{Donostia International Physics Center, Donostia-San Sebastian, Spain}
\affiliation{Fisika Aplikatua Saila, Gipuzkoako Ingeniaritza Eskola, University of the Basque Country (UPV/EHU), Donostia-San Sebastian, Spain}
\author{Maia G. Vergniory}\affiliation{Max Planck Institute for Chemical Physics of Solids, Dresden, Germany}
\affiliation{Donostia International Physics Center, Donostia-San Sebastian, Spain}
\author{Steffen Wiedmann}\affiliation{High Field Magnet Laboratory (HFML - EMFL), Radboud University, Toernooiveld 7, 6525 ED Nijmegen, The Netherlands}
\affiliation{Radboud University, Institute for Molecules and Materials, Nijmegen 6525 AJ, Netherlands}
\author{Claudia Felser}\affiliation{Max Planck Institute for Chemical Physics of Solids, Dresden, Germany}
\author{Philip J. W. Moll${}^{\dagger}$}\affiliation{Max Planck Institute for the Structure and Dynamics of Matter, Hamburg, Germany}

\date{\today}

\maketitle
\beginsupplement
	
\subsection*{Crystal synthesis and characterization}
 CsV$_3$Sb$_5$ crystallizes in the hexagonal structure with P6/mmm space group. Following the crystal growth procedure described in Ref.\cite{ortiz2020cs}, we obtained plate-like single crystals with typical dimensions of 2 × 2 × 0.04 mm$^3$. The crystals of KV$_3$Sb$_5$ were grown by the self-flux method. K, V, and Sb with atomic ratio of 7: 3: 14 were loaded in an alumina crucible and then sealed in a tantalum tube. The sample was heated to 1000 ℃, annealed for 20 hours, and cooled down to 400 ℃ with a rate of 3 ℃/h. After that, the sample was naturally cooled down to room temperature by turning off the furnace. Hexagonal crystals of KV$_3$Sb$_5$ were obtained by dissolving the flux by water. 
 
The fabrication procedure of the membrane-based device is described in Ref.\cite{Guo2022}. The device is firstly calibrated in a commercial PPMS system with 9~T superconducting magnet for the temperature dependence of resistivity and the angular dependence of magnetoresistance. High-field magnetotransport was performed inside a 35 T Bitter magnet at the High Field Magnet Laboratory. This was done using a probe with an in-situ rotatable stage equipped with the electrical connections for magnetotransport which were read out via standard lock-in techniques (SR830 and SR860).

\subsection*{Angular dependence of magnetoresistance}
The angular dependence of magnetoresistance is measured with a rotation from c to a'-axis for both CsV$_3$Sb$_5$ and KV$_3$Sb$_5$ (Fig. S1). Here $\theta$ stands for the angle between c-axis and field direction, and a' is defined as the in-plane direction perpendicular to a-axis. For the longitudinal configuration with the field applied approximately parallel to the current direction, a negative magnetoresistance (MR) is observed in CsV$_3$Sb$_5$. This is possibly due to the reduction of boundary scattering in magnetic field, as commonly seen in clean metals where the transport mean free path is comparable to the size of the microstructure. For KV$_3$Sb$_5$, the longitudinal magnetoresistance also tends to saturate at high magnetic field yet no negative MR is observed. This suggests an enhanced scattering rate in KV$_3$Sb$_5$ likely due to K-vacancies. Therefore, the quantum oscillation amplitude is also smaller in K- compared to Cs-compound, yet clear quantum oscillations can still be observed up to 75 deg for both materials.

\subsection*{FFT analysis of quantum oscillations}
By subtracting the third-polynomial function as a MR background, we have obtained the SdH oscillations for various angles. After having identified all peaks in the Fast Fourier Transform analysis with the field window of 5 to 35 T (Fig. S2), the full angular dependence of the high oscillation frequencies ($F$ $>$ 500 T) is presented in Fig. 2. 

\subsection*{Heating-induced extrinsic second harmonic generation in KV$_3$Sb$_5$}
To verify the possible extrinsic origin of the second harmonic voltage generation, we have measured the device at the same temperature and field configuration yet with different thermal conditions (Fig. S3). By adding a slight amount of He$_4$ exchange gas to the sample chamber, the thermal link between the microstructure and the sample space is enhanced. If the second harmonic voltage we measured is completely intrinsic, this change of thermal condition shall not alter the signal at all\cite{Guo2022}. However, a change of the signal can be observed which suggests the existence of extrinsic second harmonic voltage, most likely due to the thermal gradient across the device. This is likely due to the contact resistance difference between the two current leads. Consistently a field-symmetric component is readily observed and the signal measured at $\pm$ 35 T is differed by about 20 \% for $\theta$ = -1 deg.

\section*{References}

\clearpage

\clearpage
\renewcommand{\figurename}{Fig.}
\setcounter{figure}{0} 
\begin{figure}
	\centering
\includegraphics[width = 0.98\linewidth]{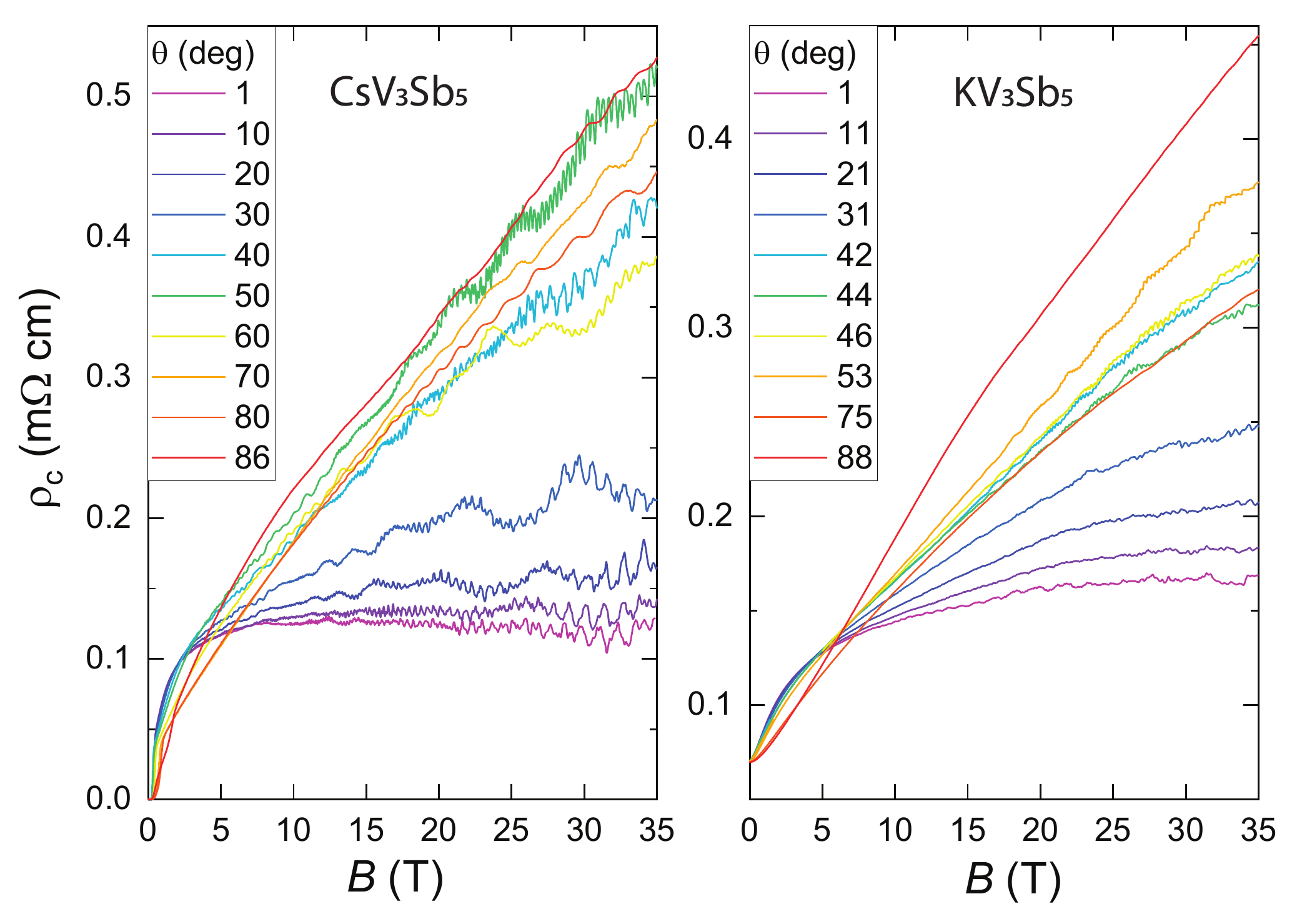}
		\caption{\textbf{Angular dependence of magnetoresistance} The angular magnetoresistance is measured with a field rotation from c to a'-axis and $\theta$ here stands for the angle between the field direction and c-axis. Clear quantum oscillations are readily observed in both materials, yet the amplitude is comparatively small for KV$_3$Sb$_5$ due to the reduced sample quality.}
	\label{SEM}
\end{figure}	
\clearpage

\begin{figure}
	\centering
\includegraphics[width = 0.9\linewidth]{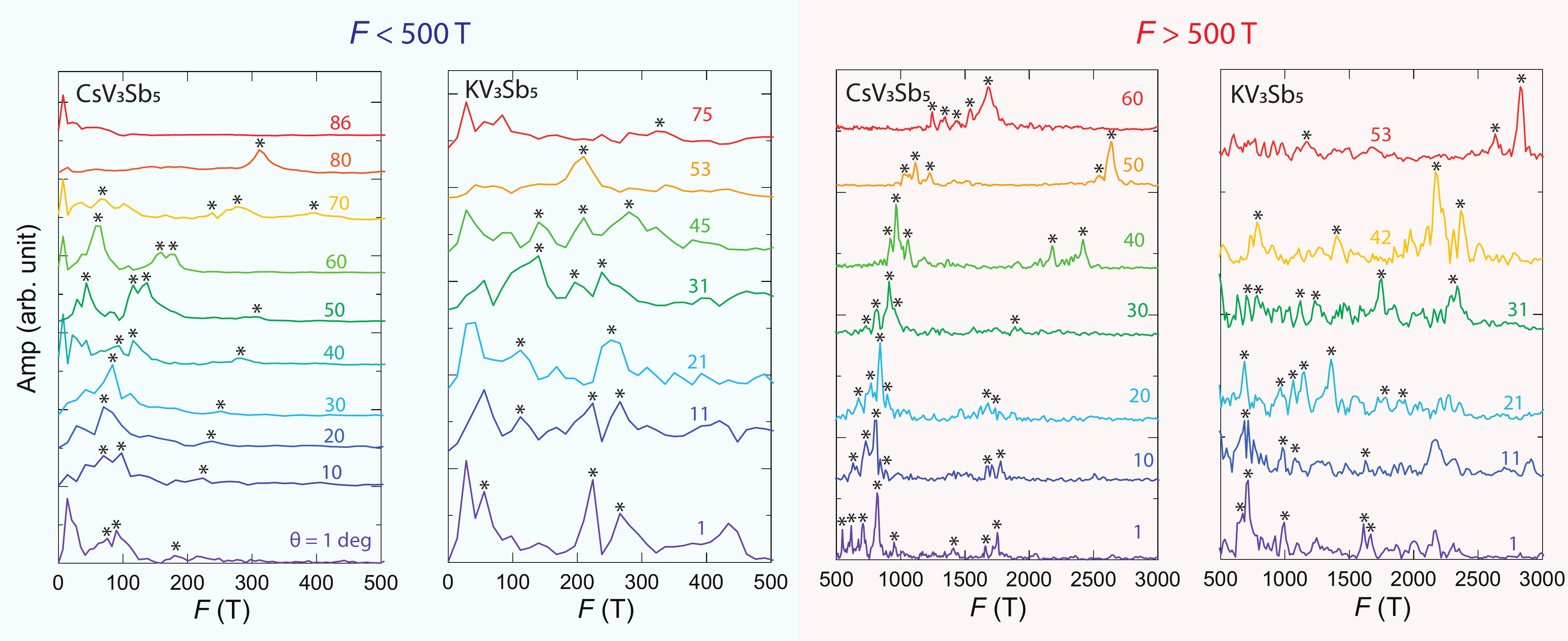}
		\caption{\textbf{Angle-dependent FFT spectrum} The Fast Fourier Transform (FFT) analysis of quantum oscillations at various angles allows us to identify the angular evolution of each peak. The corresponding frequencies are summarized in Fig. 2.} 
	\label{basicR}
\end{figure}

\begin{figure}
	\centering
\includegraphics[width = 0.9\linewidth]{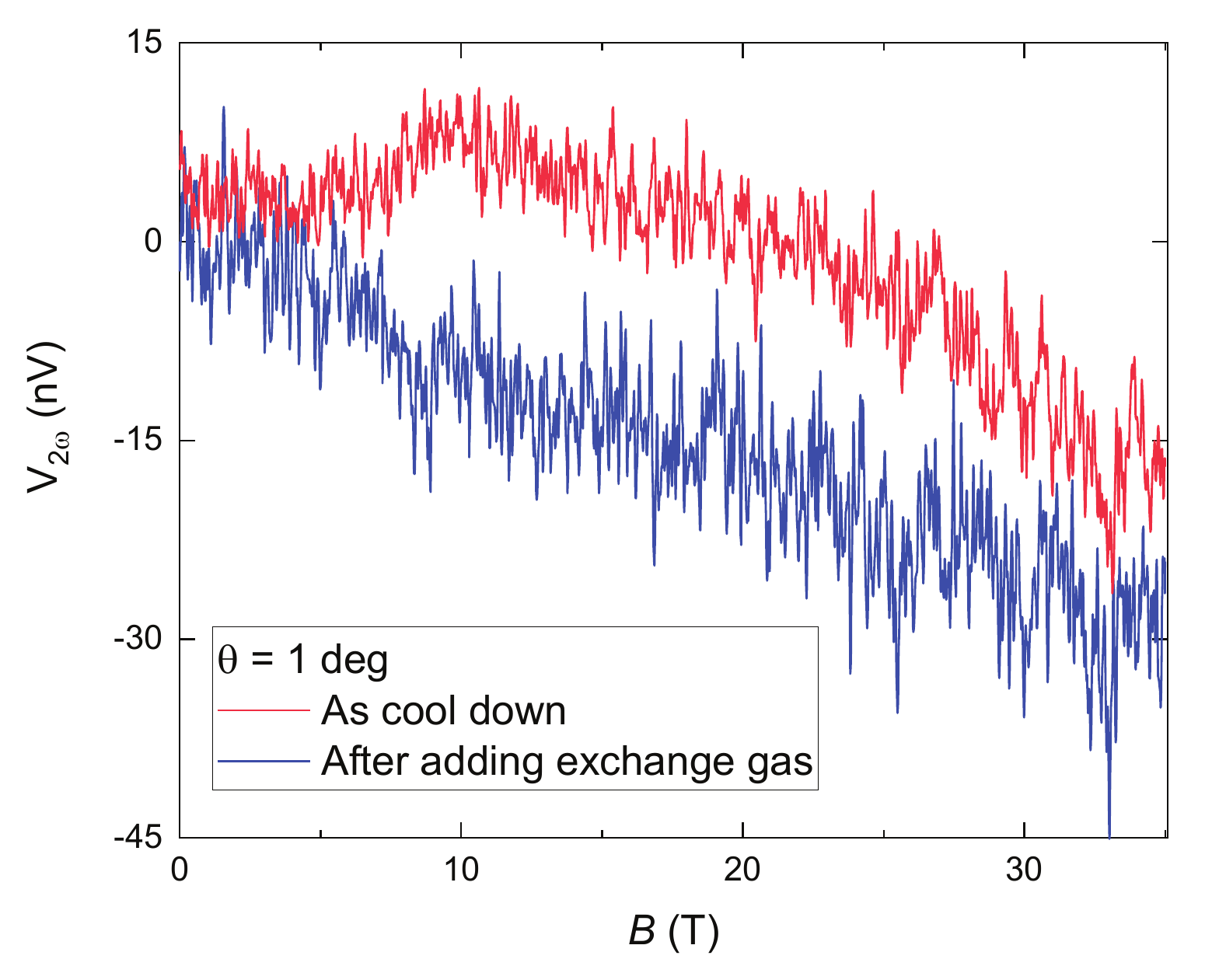}
		\caption{\textbf{Origin of extrinsic second harmonic voltage generation} The field-dependence of V$_{2\omega}$ has been measured with two different thermal conditions. By adding more exchange gas to the sample space, the signal is significantly altered, suggesting Joule heating as the origin of the observed second harmonic signal.} 
	\label{basicR}
\end{figure}